\documentclass[12pt,a4paper]{article}
\usepackage{ucs}
\usepackage[active]{srcltx}
\usepackage[utf8x]{inputenc}
\usepackage[english]{babel}
\usepackage[T1]{fontenc}
\usepackage{ae,aecompl}
\usepackage[numbers,comma,square,compress]{natbib}
\usepackage{amsmath,amssymb,amsfonts}
\usepackage{braket}
\usepackage{ifpdf}
\ifpdf
  \usepackage[pdftex,a4paper,hmargin=25mm,vmargin=25mm]{geometry}
  \usepackage[pdftex]{hyperref}
\else
  \usepackage[dvips,a4paper,hmargin=25mm,vmargin=25mm]{geometry}
  \usepackage[dvips]{hyperref}
\fi

\newcommand{\email}[1]{\href{mailto:#1}{#1}}

\newcommand{\be}{\begin{equation}}
\newcommand{\ee}{\end{equation}}
\newcommand{\bea}{\begin{eqnarray}}
\newcommand{\eea}{\end{eqnarray}}
\newcommand{\nn}{\nonumber}

\newcommand{\mr}[1]{\mathrm{#1}}
\newcommand{\cS}{\mathcal{S}}
\newcommand{\cN}{\mathcal{N}}
\newcommand{\cE}{\mathcal{E}}
\newcommand{\lP}{l_\mathrm{P}}
\newcommand{\mP}{m_\mathrm{P}}

\begin{document}

\begin{center}
{\bf\Large On gravity as an entropic force}\\
\vspace{1.5em}
{\bf M. Chaichian, M. Oksanen and A. Tureanu} 
\footnote{Electronic addresses: \email{masud.chaichian@helsinki.fi}, \email{markku.oksanen@helsinki.fi},\\  \email{anca.tureanu@helsinki.fi}}\\
\vspace{1em}
\textit{Department of Physics, University of Helsinki\\
 P.O. Box 64, 00014 Helsinki, Finland}
\end{center}

\vspace{1.5em}
\begin{abstract}
We consider E. Verlinde's proposal that gravity is an entropic force -- we shall call this theory entropic gravity (EG) -- and reanalyze a recent claim that this theory is in contradiction with the observation of the gravitationally-bound ground state of neutrons in the GRANIT experiment. We find that EG does not necessarily contradict the existence of gravitationally-bound quantum states of neutrons in the Earth's gravitational field, since EG is equivalent to Newtonian gravity in this case. However, certain transitions between the gravitationally-bound quantum states of neutrons, in particular spontaneous decays of excited states, which can hopefully be observed in future experiments, cannot be explained in the framework of EG, unless essential ingredients are introduced into it. Otherwise, a quantized description of gravity will be required.
\end{abstract}

\vspace{2em}

It is indeed the case that (time-independent) Newtonian gravity (NG) can be expressed as an entropic force, what has been proposed to follow from thermodynamics of holographic screens \cite{Verlinde}. We shall call this theory entropic gravity (EG). In fact the two theories, NG and EG, have been shown to be equivalent \cite{H}: NG$\,\Leftrightarrow\,$EG. Moreover, some of the assumptions and interpretations made in Ref.~\cite{Verlinde} are unnecessary \cite{H} and it is still unclear whether we can interpret the ``temperature'' and  the ``entropy'' of space (or spacetime) in the usual sense. Currently, EG appears to be a reformulation of NG, not a replacement nor an explanation of its origin. The case of general relativity (GR) is more involved because the derivation of GR from EG still requires one to assume several facts about GR: e.g. the generalized gravitational potential, with matter being encoded by the covariantly conserved energy-momentum tensor, and equivalence principle for local inertial frames. Hence, it is possible that EG could fail to reproduce GR. It is, however, conceivable that this problem could be overcome in the future.
Thus, it would be premature to conclude that gravity has a purely thermodynamic origin or even a fundamentally holographic character. Still, NG and GR may happen to admit such a thermodynamic reformulation. Such a holographic reformulation could turn out to be useful if it could be applied to a system that cannot be handled in the conventional formulation of gravity -- that is when bulk degrees of freedom do not exist. In all conventional (nonrelativistic) situations EG should agree with NG since it implies NG and vice versa.

The GRANIT experiment \cite{N} uses ultra-cold neutrons and a horizontal neutron mirror in the Earth's gravitational field for the realization of the gravitational quantum well (GQW), also known as the quantum bouncer. A horizontal neutron absorber is placed above the mirror in order to remove neutrons with too large vertical momentum -- states that would overlap with the absorber. This setup has enabled the experimental study of quantized gravitationally-bound states of neutrons for the first time.
The quantum stationary states of neutrons are described by the one-dimensional (vertical coordinate $z$) Schr\"odinger equation with the GQW potential. 
The ground state is predicted to have an energy $1.4\,\mr{peV}$ and a vertical extent $z_1=13.7\,\mr{\mu m}$. When the height $h$ of the slit between the neutron mirror and the absorber is smaller than $z_1$, neutrons do not fit in the slit and therefore no neutron transmission through the slit can occur. The data show that the neutron transmission rate $n(h)$ increases at the absorber height $h\approx 15\,\mr{\mu m}$ in a stepwise manner. This agrees well with the quantum mechanical prediction, which in turn differs greatly from the classical prediction $n(h)\sim h^{3/2}$.
Using a position-sensitive neutron detector for improved accuracy in Ref.~\cite{N2}, the first and second quantum states have been observed and the measured sizes of the neutron wavefunctions, $z_1^\mr{exp}= 12.2 \pm 1.8_\mr{syst} \pm 0.7_\mr{stat}\,\mr{\mu m}$ and $z_2^\mr{exp}= 21.6 \pm 2.2_\mr{syst} \pm 0.7_\mr{stat}\,\mr{\mu m}$, agree with the predicted values.
There is little reason to doubt the existence of the third and higher energy states.
This is a convincing evidence for the existence of gravitationally-bound quantum states of neutrons in the GQW formed by the Earth's gravitational field and the mirror.
For further analysis of the experiment, see also Refs.~\cite{Voronin,W,N3}. A method for observing magnetically-induced resonance transitions between gravitationally-bound quantum states of neutrons in the GRANIT spectrometer has been presented in Ref.~\cite{Kreuz} (this paper also includes an extensive list of references regarding the original experiment). Such an experiment could provide precise measurements also for the higher energy levels.

We should note that the experiment \cite{N} does not exhibit any clear quantum gravity effects. It is the quantum (wave) nature of matter that provides the quantum effects in this experiment, while gravity is included via the classical potential of NG.
This experiment proves that the classical description of gravity is sufficient at the scale of a nucleon, and certainly at the neutron wavelength scale of some microns, as long as time-independent quantum mechanics (QM) is considered.

Since EG and NG are equivalent theories and the microscopic origin of gravity is not relevant to the experiment \cite{N}, we would expect that EG agrees with the result of this experiment exactly the way NG does.
Therefore, the conclusion of Ref. \cite{Kobakhidze} that EG and NG produce different predictions in the case of the GRANIT experiment \cite{N} deserves a more in-depth study. Our arguments show that the contradiction between EG and NG claimed in  \cite{Kobakhidze} is unfounded, since the quantum mechanical description of the experiment only assumes the presence of a classical uniform time-independent gravitational potential, and EG is known to produce the same scalar potential $\Phi$ of NG for generic matter configurations \cite{H}.
The crucial part in the analysis of \cite{Kobakhidze} is the treatment of neutron states in EG. Generally speaking, according to Ref.~\cite{Kobakhidze}, pure quantum mechanical states cannot extend in the direction of the emergent coordinate of EG, say $z$, which points in the direction of the entropy gradient, and hence the translation operator for the $z$-direction must be nonunitary.
In reality, the exact nature of the relation between EG and QM is very difficult to study because the microscopic origin of EG -- presumably quantum mechanical description of holographic screens -- is utterly unknown. The analysis of \cite{Kobakhidze} and in particular the problems with it, are discussed next.

In order to apply the entropy argument of EG to the neutron states of the GRANIT experiment, some nontrivial assumptions about the microscopic origin of EG were made in Ref.~\cite{Kobakhidze}.
First, consider the density operator of a holographic screen $\cS(z)$ \cite{Kobakhidze}:
\be\label{EGQM1}
\rho_\cS(z) = \sum_{i(z)} p_{i(z)} \ket{i(z)}\bra{i(z)}
\quad ;\ i(z)=1,2,\ldots,N(\cE(z),z)\,,
\ee
where $z$ is the emergent vertical coordinate, $\cE(z)=Mc^2+mgz$ is the total energy and $N(\cE(z),z)$ is the number of microstates or ``bits`` in $\cS(z)$. All the $N$ microstates $\ket{i(z)}$ are assumed equally probable, $p_{i(z)}=1/N$ (microcanonical ensemble), and hence the entropy is maximal. For a black hole horizon this is certainly a widely accepted property. But one can wonder whether it can hold for every holographic screen. If $\cS$ already has the maximal entropy, how can it encode everything inside $\cS$ -- including gravity -- that is supposed to be an entropic force? Apparently, nothing happens inside an isolated black hole in the macroscopical sense, but that is not the case for most parts of the Universe.

The state of the screen-neutron system is defined in the spirit of EG as follows: when $|\Delta z| < \lambda=\frac{\hbar}{mc}\approx 0.21\,\mr{fm}$ (Compton wavelength), the neutron is described by a fragment $\cN$ of the screen $\cS$, and when $|\Delta z| > \lambda$ the total state is the tensor product $\rho_\cN(z+\Delta z) \otimes \rho_{\cS/\cN}(z)$, where $\cS/\cN$ denotes the screen without the fragment $\cN$. Then, the key assumption [right above Eq. (9) in \cite{Kobakhidze}] is that for $|\Delta z| > \lambda$
\be\label{EGQM2}
\rho_\cS(z+\Delta z) = \rho_\cN(z+\Delta z) \otimes \rho_{\cS/\cN}(z) \,.
\ee
As a result of \eqref{EGQM2}, the entropy gradient is solely associated with the fragment $\cN$ which describes the neutron [see Eq. (10) in \cite{Kobakhidze}]. It is then claimed \cite{Kobakhidze} that this is a basic result of Ref.~\cite{Verlinde}. This is not the case -- the entropy gradient is supposed to be associated with the screen $\cS(z)$ (or a sufficiently large part of the screen) that encodes more information than just the fragment $\cN$. Indeed, the assumption \eqref{EGQM2} implies that in the absence of the neutron the density operator is the same at all heights $z$, $\rho_{\cS/\cN}(z+\Delta z) = \rho_{\cS/\cN}(z)$, and consequently there is no entropy gradient. However, a nonzero entropy gradient must exist regardless whether there is a test particle present (neutron in our case), because the entropy of the screen $\cS(z)$ increases monotonically with $z$. This can be seen, for example, by defining the entropy of EG and the potential $\Phi$ of NG in terms of each other as in Ref.~\cite{H}. Thus the assumption \eqref{EGQM2} is not correct.

Another way to see why Eq.~\eqref{EGQM2} is incorrect, is to analyze the microstates of the screen $\cS$ in Eq.~\eqref{EGQM1}.
In order to see how the number of microstates $N$ depends on $z$, consider that the Earth is spherical and the holographic screens $\cS(z)$ around it are 2-spheres with radius $R+z$, where $z\ll R\approx 6370\,\mr{km}$. Then we obtain
\bea\label{EGQM3}
N(z) &=& \frac{\text{Area of }\cS(z)}{G} = \frac{4\pi(R+z)^2}{\lP^2} =  N(0) + 4\pi\lP^{-2}(2R+z)z \,,\nn\\
\Delta N &\equiv& N(z+\Delta z)-N(z) = 4\pi\left[2(R+z)\Delta z+(\Delta z)^2\right] \lP^{-2} \gtrsim 4\pi \times 10^{61} \,,
\eea
where $G$ is the gravitational constant and $\lP$ is the Planck length.
The number of microstates in $\cS(z+\Delta z)$ is immensely larger than in $\cS(z)$, even if we consider only a small part of the screen that is sufficient for the experiment. Thus, in Eq.~\eqref{EGQM2} the density operator in the left-hand side consists of many more microstates than the one in the right-hand side.

Therefore, we are bound to conclude that one cannot treat neutron states in EG in the simple way it is done in \cite{Kobakhidze}.
Unfortunately, this problem in \cite{Kobakhidze} is not easy to correct. In order to correct Eq.~\eqref{EGQM2}, one needs to find a relation for the density operators of the screens $S(z)$ at different $z$. As noted above, the spaces of states for these screens have vastly different dimensions. Such a relation may indeed require us to first understand the nature of the microscopical degrees of freedom of holographic screens, the piece of knowledge we do not have.

So, how do we treat EG in the cases where the quantum nature of matter is relevant?
Is it sufficient to include the effect of EG into nonrelativistic QM of a ultra-cold neutron in the form of  effective gravitational potential produced by EG?
Indeed, EG appears to be especially ill-suited for application to single particle QM because of the lack of common concepts -- neither the temperature nor the entropy of EG makes sense in single particle QM. So in such cases could we simply opt to use NG instead, since it is equivalent to EG? In this case EG and NG would of course produce identical results.
Or do we need to take into account the microstructure of holographic screens, when we consider the neutron states with EG, in particular, when we consider states extending in the emergent dimension?

It is reasonable to assume that the microstructure of EG becomes relevant at the similar length scale as quantum gravity effects are expected to become dominant ($\sim\lP$), since each microscopic degree of freedom on a screen occupies a Planck cell.
In EG gravity is a thermodynamic (statistical) effect, which exists only at length scales much greater than $\lP$, while conventionally we think that gravity is an effective result of a fundamental quantal interaction. At length scales greater than $\lP$, these two descriptions of gravity can indeed be indistinguishable. In the time-independent nonrelativistic setting, this is really the case.
The fact that the Compton wavelength of a neutron ($\sim 1\,\mr{fm}$) and the characteristic length scale of the experiment \cite{N} ($\sim 1\,\mu\mr{m}$) are far above the length scale of quantum gravity, justifies the use of the Newtonian gravitational potential in the description of the GRANIT experiment, which is confirmed by the data. Therefore, EG does not necessarily contradict the existence of the gravitationally-bound eigenstates in the GRANIT experiment any more than NG does.

However, when we start to consider the dynamics of the gravitationally-bound states in GQW,  things change. As an example and an analogy, let us consider electromagnetism and in particular the bound states of the hydrogen atom. In the (time-independent) description of the energy eigenstates the electromagnetic field need not be quantized. Even the absorption of radiation and the radiation-induced emission can be understood using the semiclassical description, where the electromagnetic field is treated classically. Understanding the phenomenon of spontaneous emission associated with the decay of an excited state of the atom requires us to quantize the electromagnetic field. A similar reasoning can be applied to the GQW realized in the GRANIT experiment. When a gravitationally-bound excited state spontaneously decays, the excess gravitational energy has to be carried away by a quantum of gravity, namely the graviton. It would be very strange, if all the transitions from the higher gravitationally-bound states to the lower ones would be prohibited in every system. In EG, seemingly there is no way to quantize gravity because it is a thermodynamic effect, not a fundamental interaction. Yet, it is not known whether EG can accommodate the notion of graviton, since graviton may also appear as an emergent concept in this theory, much like that in AdS/CFT duality or as phonon in solid state physics. If EG will eventually turn out unable to contain graviton as an emergent notion, the existence of a spontaneous decay of a gravitationally-bound state would falsify EG, as any other observation of the graviton spectrum would do.
In the GRANIT experiment, however, the transition rates for spontaneous decays are too low to be observable \cite{P}; for example the transition rate from the first excited state to the ground state is of the order of $10^{-77}\,\mr{s}^{-1}$.
Still, the emission of a graviton from a gravitationally-bound state could be a more viable way to indicate the existence of the graviton, compared to a direct detection which is practically impossible (see \cite{R} and references therein).

As mentioned above, in the relativistic case there are problems with the derivation of GR from EG. Moreover, it has been shown \cite{L} that a region undergoing accelerated expansion (inflation) must have a negative temperature, which is very problematic. Thus, EG may fail also with respect to its relation to GR, not just on the quantum gravity ground.

As a summary, although the analysis of \cite{Kobakhidze} is not fully consistent, EG as an explanation for the origin of gravity still needs further studies for its justification, what would provide conceptual challenges in the understanding of the graviton problem. Since gravitationally-bound quantum states of neutrons in the Earth's gravitational field have already been observed, we have every reason to expect that eventually a quantized description of gravity will become necessary. Thus we suspect that the thermodynamic argument alone is not entirely sufficient to explain gravity even at relatively large distances. This does not necessarily mean that gravity cannot be fundamentally holographic or emergent -- it is entirely possible that space-time and (quantum) gravity could emerge from some underlying theory.

In loop quantum gravity the holographic principle and the argument of Verlinde have been used to derive Newton's law of gravity in an appropriate limit \cite{S}. This gives some hope that, eventually, it may be possible to accommodate the existence of graviton in EG as an emergent quantum of gravity much like that in AdS/CFT duality or as phonon in solid state physics.

An interesting prospect is the possibility to combine the ideas of EG and the classicalization of gravitons \cite{D1,D2,D3,D4}. In the classicalization phenomenon, an interaction with a center-of-mass energy higher than the Planck mass ($\sqrt{s}>\mP$) produces a classical black hole configuration -- a classicalon -- which may self-unitarize gravity at high energies. The entropy of the black hole has a precursor given by the number of soft quanta composing the classicalon, and it could even be described in terms of classicalon states.
Such an entropy could quantitatively account for gravity in the EG approach.

\paragraph{Acknowledgements}
We are deeply grateful to Gia Dvali for several illuminating discussions and suggestions.
We also wish to thank Archil Kobakhidze for critical comments and Valery Nesvizhevsky for useful discussions on the GRANIT experiment.
 M. O. is supported by the Jenny and Antti Wihuri Foundation. 
The support of the Academy of Finland under the Project Nos. 136539 and 140886 is gratefully acknowledged.

\end{document}